\documentstyle[prl,aps,floats,epsf,epsfig]{revtex}

\begin{document}

\draft

\title{Planar and Stripe Orders of Doped Mott 
Insulators In Dual Spin Models}

\author{Marco Bosch and Zohar Nussinov}
\address{Institute Lorentz for Theoretical Physics, Leiden University\\
P.O.B. 9506, 2300 RA Leiden, The Netherlands}
\date{\today  
; E-mail:mbosch@lorentz.leidenuniv.nl, zohar@lorentz.leidenuniv.nl
}

\twocolumn[

\widetext
\begin{@twocolumnfalse}

\maketitle
\begin{abstract}

We focus on very general, very large U, doped Mott Insulators
with arbitrary hopping and interactions.
We provide simple testimony to the competition between 
magnetic and superconducting orders
in these systems. By mapping 
hard core bosons, spinless, and spinful fermions onto XXZ models, 
we aim to make very simple precise statements. We try to address  
optimal and expected filling fractions of 
holes within the plane and on stripes 
in a variety of hole and hole pair geometries. 
We examine the role of attractions/repulsion
amongst hole pairs and single holes,
and provide trivial expected numerical values 
for filling fractions in various 
scenarios. We demnostrate that plaquette states seem to 
naturally provide the correct 
stripe filling fractions.

\end{abstract}


\vspace{0.5cm}

\narrowtext

\end{@twocolumnfalse}
]

\section{Introduction and Outline}

Throughout this paper we aim to address questions concerning
planar and linear (stripe) filling fractions of doped Mott insulators
by trying to see how much may be gleaned by mapping such 
systems onto spin models. The answers that we find depend 
greatly on the the assumed form of our constituent 
particles (single holes, pairs) and their various 
geometries (plaquettes, rungs, bonds). We 
will demonstrate that plaquette states seem to 
naturally provide the correct 
stripe filling fractions.
 By mapping onto spin models, we will also be able to
examine various competing orders.
Amongst other things, we will demonstrate 
(not unexpectedly) that in the limit
of large on-site repulsions $U$, magnetic and 
superconducting orders always compete.

The outline of the paper is as follows.
In Section(\ref{Standard-Model}), we 
lay out the general standard model 
of Doped Mott Insulator which 
will form the focus of our discussion.
In Section(\ref{FIP}), we examine the 
problem of spinless and spinful particles 
in the plane by a mapping to a spin model.
We will see, in a special
yet general set of models
another rigorous example of the competition 
between magnetic
and superconducting orders.
Within these spin models that result
for large U, we illustrate how magnetic order
as well as the physically transparent 
number order (both portrayed by $S_{z}$
in two different spin representation) compete  
with the dual superconducting order
encapsulated by bilinears in the planar
$\vec{S}_{\perp}$ multiplied
by exponentials in of a topological nature.

In Section(\ref{PLAQUETTE}), we map 
a model of hard core bosons (hole pairs) in non-overlapping plaquettes 
with arbitrary finite range interactions and hoppings
onto a planar XXZ model. We find that if the physics is 
indeed dominated by such attractive pair states then, within the 
ground state, the average hole occupancy
per site within the optimal doped state 
may be 1/4. The large deviation from
the observed optimal doping in
most doped Mott insulators is 
hardly surprising and points
to the inadequacies of looking
at attractive plaquette pairs alone 
within the plane: not unexpectedly,
an analysis of the cuprates 
based solely on notions of
doped Mott insulators and concurrently 
assuming only Bose pairs of the 
plaquette size might be flawed.

We next examine, in Section(\ref{plaq_on_stripe}),
plaquette pairs on bond-centered stripes and find, under a variety
of circumstances, near 1/4-filling. We discuss the possibility of
phase separation or modulation
of the hole pairs along the bond centered stripe.
When we relax the condition that
diagonal pairs do not have to be 
in non-overlapping plaquettes, we will find,
in the large hopping amplitude ($t$) limit, a 1/3-filling 
fraction of stripes. Higher and similar filling
fractions are found for pairs on single rungs
and legs. From this 
simple exercise we conclude
that bond centered stripes cannot, perhaps, 
be described by simple pairs roaming 
the stripe axis, and that 
if pairs indeed do dominate
the asymptotic low energy stripe scale 
physics as we suggested in an earlier
paper \cite{us}, then they must be effectively 
confined to non-overlapping plaquettes in accord with
pictures suggested by the DMRG calculations of White and 
Scalapino \cite{WS}. 

We the proceed with a similar analysis
of fermions on bond centered stripes (Section(\ref{single})), 
and find that these lend themselves 
to near 1/4 filling. Our description
fortifies earlier work by 
Nayak and Wilczek \cite{nayak}.

\section{The Model}
\label{Standard-Model}

We will focus our attention on
the relatively standard model of
doped Mott insulator, the 
extended Hubbard model:

\begin{eqnarray}
\label{Hubbard model+}
H=- \sum _{\langle ij\rangle \sigma } t_{ij} (c_{i\sigma }^{\dag }c_{j\sigma
}+c_{i\sigma }^{\dag }c_{j\sigma })+\sum _{i}Un_{i\uparrow
}n_{i\downarrow} \nonumber
\\ + \sum_{ij} V_{ij} n_{i} n_{j},
\end{eqnarray}%
where \( c_{i\sigma }^{\dag } \) creates an electron on site \( i \)
with spin \( \sigma  \) and \( j \) is a nearest neighbor of \( i \).
This model contains both the movement of the electrons (hopping) (\( t \),
kinetic energy) and the interactions of the electrons if they are
on the same site (\( U \), potential energy).  The Mott
insulating nature is captured by this on-site repulsion
which greatly inhibits hole motion. We have
added an additional term representing
all possible number-number interactions
(of all ranges); these may be result from myriad interactions-
e.g. Coulomb repulsions, interactions mediated by phonons.
The number occupancy
\begin{eqnarray}
n_{i \sigma} = c_{i \sigma}^{\dagger} c_{i \sigma} \nonumber
\\
n_{i} = \sum_{\sigma} n_{i \sigma}.
\end{eqnarray}

As well known, 
in the infinite $U$ limit, trivial spin-charge 
separation occurs in any dimension. 
The charge degrees of freedom may
be trivially encapsulated by spinless degrees
of freedom. For a review of this
principle, the reader is invited
to read Appendix(\ref{any}). 

At large $U$, the extended Hubbard model
of Eqn.(\ref{Hubbard model+}) may be
related to an extended t-J model,
\begin{eqnarray}
H =
- \sum_{\langle i j \rangle, \sigma}  t_{ij} (c_{i, \sigma}^{\dagger} c_{j, \sigma}
+ H.c.) +  \sum_{\langle i j \rangle}  J_{ij} \vec{S}_{i} \cdot
\vec{S}_{j} \nonumber
\\ + \sum_{ij} V_{ij} n_{i} n_{j},
\label{tJ}
\end{eqnarray}
where $\vec{S}_{i} = \sum_{\sigma \sigma^{\prime}} c_{i
\sigma}^{\dagger} \vec{\sigma}_{\sigma, \sigma^{\prime}} c_{i,
\sigma^{\prime}}$
is the spin of the electron at site $i$,  
$\vec{\sigma}$ are the Pauli matrices, and there is a constraint
of no double occupancy of any site $i$ ($n_{i} = \sum_{\sigma}
c_{i,\sigma}^{\dagger}c_{i,\sigma}$ has expectation values  0 or 1).
The reduction to a Hilbert space where no doubly
occupied sites occur (the Gutzwiller projection)
will be automatically incorporated in 
all things to come.

\section{Competing Orders For Fermions In The Plane}
\label{FIP}

\subsection{Spinless Fermions- A Competition
Between Charge And Superconducting Orders}

With possible applications to the limit of infinite 
on-site repulsion (U) in mind, we now examine
the competition of charge and superconducting 
order within the plane. All that we will
detail below can be derived 
straightforwardly. We take a 
slightly longer route in order to
hightlight the simple similarities
between spin and charge 
when looked at through
the prism of the Jordan-Wigner
representation of S=1 and S=1/2 
problems respectively. Although not
of any use for most practical 
applications, a high dimensional 
Jordan Wigner transformation 
has been devised by Fradkin \cite{Fradkin},
and later extended by Eliezer and Semenoff \cite{eliezer}.
Very nice novel extensions to various spinful
cases were recently advanced by Batista and Ortiz \cite{batista}.
The basic message is that the classic Jordan-Wigner \cite{Jordan} 
transformation rigidly linking spinless fermions and S=1/2 spins
in one dimension can, quite naturally, be
extended to higher dimensions. The only complication
is that now the kink operators that code 
for the statistics transmutations become 
high dimensional topological objects. 
To be more precise, the standard one dimensional string
operator appearing in the usual Jordan-Wigner 
transformation is replaced by its more
general counterpart
\begin{eqnarray}
K_{j} =  \exp[i \sum_{\vec{k}} \theta(\vec{k},\vec{j}) n_{\vec{k}}],
\end{eqnarray}
with $\theta(\vec{k},\vec{j})$ the angle between $(\vec{k}-\vec{j})$
and a fixed ray (in the one dimensional case the 
``angle of site'' $\theta$ reduces to either $\pi$ (for $j<k$) or zero
(for $k \ge j$) leading to the standard string operator). 
In terms of these new kink operators, 
the Jordan-Wigner transformation then reads
\begin{eqnarray}
S_{j}^{+} = c_{j}^{\dagger} K_{j}, \nonumber
\\ S_{j}^{-} = K_{j}^{\dagger} c_{j}, \nonumber
\\ S_{j}^{z} = n_{j} - 1/2.
\end{eqnarray}
The simple pair operator
$\Delta_{ij} \equiv c_{i}^{\dagger} c_{j}^{\dagger}$
may be expressed in terms of the spin
variables. In the aftermath, we find 
that the superconducting 
pairing operator may be 
expressed as a product of kink 
variables with the XY components
of the spins: $S^{\pm}_{i}$ and $S^{\pm}_{j}$.
The incompatibility 
of charge ($n$ or $S_{z}$) and superconducting 
phase (descendant from $ \vec{S}_{\perp}$)
orders is trivially reflected from the 
non-vanishing commutator
\begin{eqnarray}
[S^{z} , S^{\pm}] = \pm S^{\pm}.
\label{abc}
\end{eqnarray}
Explicitly, inverting all matters to
the spin representation,
\begin{eqnarray}
c_{i}^{\dagger} = S_{i}^{+} K_{i}^{-1} =   
S_{i}^{+} \exp[- i\sum_{\vec{r}} \theta(\vec{r},\vec{i}) S^{z}_{r}],
\end{eqnarray} 
the commutator
\begin{eqnarray}
[n_{k},\Delta_{ij}] = 
(\delta_{ik} + \delta_{jk}) \Delta_{ij},
\label{usual}
\end{eqnarray}
which is compatible with $\Delta_{ij}= \exp[i (\phi^{C}_{i}+ \phi^{C}_{j})]$,
and $n_{k} = - i \frac{\partial}{\partial \phi^{C}_{k}}$
with $C$ denoting charge.

If Eqn.(\ref{Hubbard model+})
captured all of the physics,
then computing optimal doping
would amount to the determination
of the value of $\langle S_{z}  \rangle$
when we aim to maximize the topological
pairing order parameter $\Delta$.

\subsection{Spinful Fermions- A Competition between Magnetic
and Superconducting Orders}

Just as in the spinless case, the simple commutation relations that we 
will soon derive can be immediately 
seen by direct computation (no Jordan-Wigner 
transformations are neccessary). Nevertheless,
in order to highlight the similiraties between
the charge and spin sectors as 
doublet (S=1/2) and triplet (S=1) 
representations of similar entites 
we will employ the Jordan-Wigner 
representation once again.
Batista and Ortiz \cite{batista} extended
the Jordan-Wigner transformations to spinful
fermions. The operator
\begin{eqnarray}
\overline{c}_{j \sigma}^{\dagger} = c_{j \sigma}^{\dagger}(1-n_{j
-\sigma}),
\end{eqnarray}
along with its conjugate, may be transformed 
into spin S=1 operators in much the same way 
as for the spinless case. We note that within the large
$U$ limit,  $\overline{c}_{j \sigma}^{\dagger} 
\to c_{j \sigma}^{\dagger}$. Any Hamiltonian 
in $\{\overline{c}_{j \sigma}^{\dagger}, \overline{c}_{j \sigma}\}$
has the Gutzwiller (no double occupancy) projection 
automatically built into it.
The general t-J Hamiltonian of Eqn.(\ref{tJ}) 
will undergo no change when expressed in terms
of $\{\overline{c}_{j \sigma}^{\dagger}, \overline{c}_{j \sigma}\}$
instead of $\{c_{j \sigma}^{\dagger}, c_{j \sigma}\}$.  

The transverse
(XY) components of the spin may be 
written as \cite{batista}
\begin{eqnarray}
S_{j}^{+} = 2^{1/2} (\overline{c}_{j, \sigma=+}^{\dagger} K_{j} +
K_{j}^{\dagger}\overline{c}_{j, \sigma=-}).
\end{eqnarray}
The z-component of
the spin transforms as the
on-site magnetization,
\begin{eqnarray}
S_{jz} = \overline{n}_{j, \sigma=+} - \overline{n}_{j,  \sigma=-},
\label{Sjz}
\end{eqnarray}
with $\overline{n}_{j \sigma} = \overline{c}_{j \sigma}^{\dagger} 
\overline{c}_{j \sigma}$. As before,
\begin{eqnarray}
K =  
\exp[i \sum_{\vec{k}} \theta(\vec{k},\vec{j}) \overline{n}_{k}] =
\exp[i \sum_{\vec{k}} \theta(\vec{k},\vec{j}) (S^{z}_{k})^{2}],
\end{eqnarray}
where the, new, second equality follows from Eqn.(\ref{Sjz})
and the inbuilt constraint of no double occupancy.
When we employ the inverse transformation, we
find that $\Delta_{ij} = c_{i \sigma}^{\dagger} 
c_{j -\sigma}^{\dagger}$ contains, similar to the case before, 
the transverse components
of the spin $S_{\pm}$ multiplying 
topological operators $K^{-1}$, which 
are simple exponentiated
products in $\{(S^{z}_{j})^{2}\}$.  
Once again the non-commuting
character of the spin components
(Eqn.(\ref{abc})) disallows concurrent
ideal magnetic and superconducting orders.
Explicitly,
\begin{eqnarray}
[S_{k}^{z},\Delta_{ij}]=  -\frac{1}{2} (\delta_{ik} -
\delta_{jk}) \Delta_{ij}. 
\label{s-c}
\end{eqnarray}
This is compatible with 
\begin{eqnarray}
S_{k}^{z} = \frac{i}{2} ~\frac{\partial}{\partial \phi^{M}_{k}},
\nonumber
\\ \Delta_{ij} = \exp[i(\phi^{M}_{i}-\phi^{M}_{j})],
\end{eqnarray}
with the superscript $M$ denoting ``Magnetic''.
Trivially,
\begin{eqnarray}
[S^{z}(\vec{q}),\Delta(\vec{p})] \neq 0,
\end{eqnarray}
with $\vec{q}$ and $\vec{p}$ arbitrary wavenumbers.

Examining Eqs.(\ref{usual},\ref{s-c}), we note 
that, apart from an important minus sign, 
magnetism competes with superconductivity
in much the same way as the charge (number) field 
competes with pairing for spinless fermions.
This is, of course, not all too surprising. 
We merely wished to highlight
the similarities between the spin 
and charge sectors as S=1 and S=1/2
representations of a similar
problem when viewed through  
the Jordan-Wigner
transformation.

All of what we detailed henceforth 
was for singlet pairing. For a hypothetical triplet 
pair operator $\Delta^{(t)}_{ij} =  c_{i \sigma}^{\dagger} 
c_{j \sigma}^{\dagger}$ the
commutation relations are trivially 
identical to those of the 
number and pairing correlations
of Eqn.(\ref{usual}),
\begin{eqnarray}
[n_{k},\Delta^{(t)}_{ij}] = 
\frac{1}{2} (\delta_{ik} + \delta_{jk}) \Delta^{(t)}_{ij}.
\label{triplet}
\end{eqnarray}

To summarize, in a well known simple 
rubric,  
\begin{eqnarray}
[\mbox{Spin}, \mbox{Pairing}] \sim [\mbox{Charge}, \mbox{Pairing}] 
 \neq 0.
\end{eqnarray}
We further reiterate that the charge and spin sectors
can be viewed as a doublet (S=1/2) and tripet (S=1)
representations of similar entities. The spin
component $S_{z}$ takes on different
roles the two different Jordan-Wigner
transformations.
This is simply yet another way of
viewing matters. The similarity 
between the two sectors is
highlighted in the SO(5) 
theory of S. C. Zhang \cite{zhang}.

\section{Planar Plaquette Pair States}
\label{PLAQUETTE}

Much of our approach henceforth was inspired 
by the beautiful work of Altman
and Auerbach \cite{assa}. Let the operator $\Box^{\dagger}$
denote the creation operator
of a hole pair on a plaquette. 
The difference between pairs in the cuprates and in conventional
BCS superconductors is that within the cuprates the pair
size is very small which allows us to 
consider a reduction to plaquettes.
This is also seemed to be supported by numerical
calculations. 

\begin{figure}
\includegraphics[width=5.2cm]{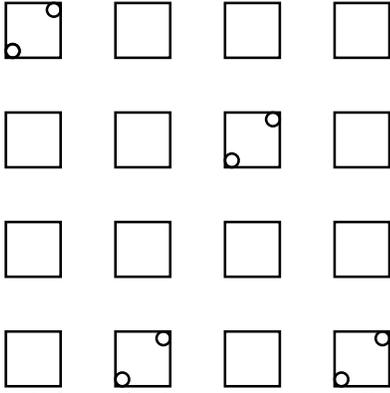}
\caption{A plaquette state. Following 
Altman and Auerbach's approach we tile the 
plane with non-overlapping
plaquettes. The small circles 
represent single holes. In the
above we consider holes 
forming pairs on plaquettes.}
\label{plaquette}
\end{figure}

\subsection{Mapping To An XXZ Model}

Employing the Matsubara-Matsuda transformation \cite{mam}
(trivially valid in any dimension) 
\begin{eqnarray}
S_{i}^{z} = n_{i} - \frac{1}{2}, \nonumber
\\ S_{i}^{+} = \Box^{\dagger}_{i}, \nonumber
\\ S_{i}^{-} = \Box_{i} ,
\end{eqnarray}
we may map the most general
Hamiltonian describing pair 
motions and interactions
of all ranges 
\begin{eqnarray}
\label{H in terms of pair operators}
\tilde{H}_{eff}= - \sum _{ \langle ij \rangle} t^{ij}
(\Box_{i}^{\dag
} \Box_{j} + \Box^{\dag 
}_{j}\Box_{i}) \nonumber
\\ + \sum_{\langle ij \rangle} V^{ij}
n_{i}  n_{j}  + \tilde{J}^{\prime} \sum_{i}
\Box_{i}^{\dag }  \Box_{i} + ...,
\label{general}
\end{eqnarray} 
to a two dimensional 
XXZ model 
\begin{eqnarray}
\! \! \! \! \! \! \! \! \! \! \! \! \tilde{H}_{eff}=
- \sum_{\langle ij \rangle} J^{\perp ij}
(S_{i}^{+} S_{j}^{-}+ S_{i}^{-} S_{j}^{+} ) \nonumber
\\ + \sum _{\langle ij \rangle} J^{z ij} 
S_{i}^{z} S_{j}^{z} - J^{\prime} \sum_{i}
(S_{i}^{z})^{2} \nonumber
\\ - \sum_{i}  h^{\prime}
S_{i}^{z}  
+ ...    
\label{sof2}
\end{eqnarray}            
with          
 $\tilde{J}^{\perp i j} = 2 t^{ij}$,
$\tilde{J}^{z i j}= V^{ij}$.

Examining the quadratic (${\cal{O}}(S^{2})$) sector for
nearest neighbors ($|i-j| = 1$), we arrive at 
the standard nearest neighbor XXZ model.
The XXZ model has two distinct phases: 
Within the Ising limit (large $|J^{z}|$) the spontaneous breaking
of $Z_{2}$ symmetry implies phase separation of the bosonic
hole pairs. By contrast, for large $|J^{\perp}|$, 
the system is in the XY limit and the ground 
state is a superfluid. At the transition
between these two phases, the model is 
$SU(2)$ symmetric. $Z_{2}$ order (number order) is dual and conjugate to 
superconducting order (XY order). We have derived 
stripes in an earlier article \cite{us} assuming weak staggered
boundary conditions on the real $S_{z}$
(a $Z_{2}$ symmetry order of yet another 
origin).

\subsection{Phase Separation or Not}

As to be expected, whether
phase separation of hole pairs 
will transpire is determined 
by the sign of the charge-charge interactions
$V_{ij}$. 

If the charge-charge interaction 
is repulsive
for all $|i-j|$ (i.e. $V_{ij} > 0$),
then in the extended spin model we will
find exchange constants $J^{z}_{ij}>0$. A positive $J^{z}$ (a repulsive
potential $V_{ij}$) favors
charge neutrality $\langle S_{tot}^{z} \rangle$
whenever the background charge is taken into
account. For an attractive potential $V_{ij} <0$
leading to $J^{z}_{ij} <0$,
there is viable ferromagnetic order. In such an
instance, the potential $V_{ij}$ 
allows (and favors) regions of different 
phase densities (i.e. phase separation). 

\subsection{Superconducting Correlations, vortices,
And A Numerical Value For Optimal Doping
For Attractive Interactions}

Whenever the expectation
value $\langle S_{z} \rangle$
vanishes we immediately
find half a pair per plaquette
or, equivalently, a doping 
of quarter of a hole per site. In order to make the
XXZ model maximally superconducting,
we would like to make 
the order as XY like 
as possible to avoid
phase separation
into bosonic hole 
pairs. If we indeed
impose from 
the outset the absence
of $Z_{2}$ like order
for the number operator
($S_{z}$) we 
find a quarter  
of a hole per site 
\begin{eqnarray}
\delta_{optimal} = 1/4. 
\end{eqnarray}
The largest $S_{z}$ (or number) fluctuations
$\langle (S_{z} - \langle S_{z} \rangle)^{2} \rangle$
occur about the symmetric point
$\langle S_{z} \rangle =0$. This feature is far more generic
than assuming a special model
of two dimensional hard core
bosons (as we have above). In high enough 
dimensions, the maximal 
fluctuations at the symmetry points
may diverge signaling a critical 
point. Large number
fluctuations enable well defined superconducting phase 
(XY) order. A trivial bound on XY order reads
\begin{eqnarray}
\langle \vec{S}_{\perp}^{2} \rangle  \le S(S+1) - \langle S_{z} \rangle^{2}.
\end{eqnarray}
As the number operator is the dual disorder operator
for the phase, in order to have maximal phase
ordering, we wish to avoid 
vortex like number expectation
values. A finite number density $(n-1/2)$,
is analogous to a finite vortex
density. This, in turn, is equivalent to
an imposed external background
magnetic flux density in a Josephson
junction array. Non-uniformities in
the charge order act as 
frustrating vortices. For maximal phase
ordering we need 
to minimize the appearance
of vortices. This entails 
lowering the effective 
magnetic flux background that 
spontaneously 
generated vortices 
must
cancel.

\bigskip

Another way of viewing matters 
at very low finite temperature (like
a low temperature classical model)
is as follows: If $J^{z}$ were zero
then adding an additional
field $h^{\prime}$ would corrupt
the XY order- the previous classical ferromagnetic
XY ground state- $\vec{S}_{\perp} = (\cos \phi,\sin \phi)$ with
the phase $\phi$ uniform over the entire plane. 
Now imagine turning on $J^{z}$ and making it negative (attractive
interactions amongst pairs).
This will only further corrupt the XY order.

The relatively large deviation of the optimal doping 
found in the cuprates from 1/4 suggests that we may
not easily view these doped Mott insulators at optimal 
or near optimal doping as merely composed
of pairs of the plaquette size
with attractive interactions.

\section{Plaquette States on Stripes}
\label{plaq_on_stripe}

The stripe constitutes a 
small system coupled to 
a large bath (the 
ambient antiferromagnet)
from which it can draw 
holes. Depending on the 
chemical potential of 
the stripe, different
hole densities will 
result. For ease we will invoke
a simplifying yet non-essential 
assumption: by virtue of the large kinetic
scale, the hopping amplitude $t$
is much larger than all potential
effects, i.e. $\sum_{j} V_{ij}$ 
and the magnetic alleviation
energy (the energy gained
by removing bad ferromagnetic
bonds along the seam of
the stripe) $J$. With
such assumptions, 
the chemical potential
trivially vanishes.
In reality, due to
the finite corrections
of both effects we expect 
the chemical potential 
$\mu/t \simeq 0$.

As we will show assuming the constituents of
the stripes to be of different geometries and nature
(e.g. hole pairs on non overlapping plaquettes, diagonal 
or parallel hole pairs) different filling fractions will
be found at the approximate particle-hole symmetric
point $\mu=0$.

We will now require that every two consecutive rungs of 
the ladder form a unique plaquette (such that these
plaquettes do not overlap)
and consider all possible pair states 
within these plaquettes. 

\begin{figure}
\includegraphics[width=8.2cm]{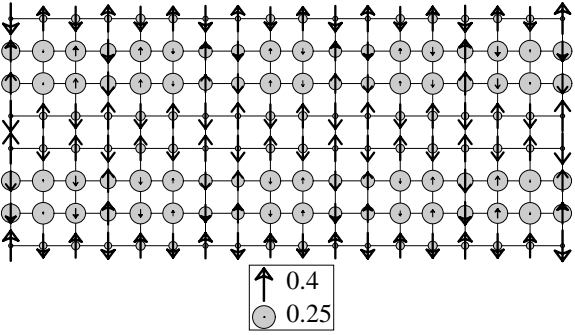}
\caption{The DMRG calculation of the $t$-$J$ model by White and
Scalapino. This and similar calculations typically employ
open or cylindrical boundary conditions.}
\label{dmrg}
\end{figure}

As we will now show, if the chemical
potential is indeed small by comparison
to the kinetic hopping amplitude $t$,
then assuming pair states
in non-overlapping plaquettes will
lead to the ``correct'' filling
fraction. By ``correct'', we allude
to the coincidence, within the large $\mu/t$
limit, with the filling fraction attained from 
an original large $U$ Hubbard 
Hamiltonian for single fermions augmented
by all possible hoppings and interactions
amongst the holes. As we show
in section(\ref{ideal}), if
we start from the original
Hubbard model and 
allow holes to do anything
that they wish from there
(e.g. remain single, pair, phase separate)
we will find that the net hole 
filling fraction is 1/4. This coincidence suggests that 
plaquette pair states are indeed viable 
candidates for the pairs states populating  
bond centered stripes.
Moreover, both numerical \cite{WS}, and
mean field \cite{marco} calculations 
lead to quarter filling as 
well as experimental data 
\cite{yamada}.

These plaquette pair states
are indeed the states observed
numerically. Moreover, as 
emphasized by \cite{assa} 
they naturally display 
$d$- wave symmetry.

As before,  we may use the operator $\Box^{\dagger}$
to denote the creation operator
of a hole pair on a plaquette
and subsequently employ the Matsubra-Matsuda transformation
with no change to obtain once again Eqn.(\ref{general}).

As seen in Fig.(\ref{dmrg}), adding a 
hole pair along the bond centered stripe
removes energetic ferromagnetic
bonds. Such an effect will
only serve as a weak sink favoring more
hole pairs to drift into the stripe.

If the charge-charge interaction, 
\begin{eqnarray}
\sum_{ij} V_{ij} n_{i} n_{j} - 2J
\sum_{i} n_{i} \nonumber
\\ =  
\sum_{ij} V_{ij} (n_{i}-1/2)  (n_{j}-1/2) + \sum_{i} n_{i} \sum_{j}
V_{ij} \nonumber
\\ - 2J
\sum_{i} n_{i},
\end{eqnarray}
is repulsive
for all $|i-j|$ (i.e. $V_{ij} > 0$),
then in the extended spin model we will
find exchange constant $J^{z}_{ij}>0$
for all $|i-j|$. A positive $J^{z}$ favors
charge neutrality $\langle S_{tot}^{z} \rangle$
when the background charge of (-1/2)  
(if $\sum_{j} V_{ij}= 2J$) is taken into
account. For an attractive potential $V_{ij} <0$
leading to $J^{z}_{ij} <0$,
there is viable ferromagnetic order. The potential $V_{ij}$ 
allows (and favors) regions of different 
phase densities (i.e. phase separation).

In the presence of a vanishing electronic chemical 
potential
\begin{eqnarray}
\mu = \sum_{j} V_{ij} - 2 J,
\end{eqnarray}
or, effectively, an external 
magnetic field $\tilde{h}$
of a similar magnitude,  
there is no spontaneous symmetry
breaking in the XXZ chain 
\begin{eqnarray}
\langle S_{z} \rangle = 0
\end{eqnarray}
and there is an average of 
one pair per two plaquettes
or a quarter of a hole
per site. This is indeed 
in accord with experimental
observations \cite{yamada}
and the numerical calculations
of White and Scalapino \cite{WS}.
In reality, the chemical
potential albeit small 
compared to $t$ is not 
zero and a small deviation
from 1/4 will be found.
Nothing is inserted by hand
here- the occupancy is 
dictated by the chemical
potential (or magnetic
field) which is finite yet may small
compared to $t$ by virtue 
of the various parameters
that it includes (i.e. $J/t$ and
$\frac{1}{t} \sum_{j} V_{ij}$).
At its very core, the unbroken
spin rotational symmetry corresponds 
to particle-hole symmetry trivially 
present at $\mu/t=0$. This symmetry
is lightly lifted by a small finite chemical
potential shift relative to its 
vanishing value the particle-hole
symmetric point.

We will later on demonstrate 
that if we assume only 
symmetrized rung states
for each of the individual
holes then no matter
how the holes interact
and whether they form
pairs, phase separate, 
or not, a bond 
centered stripe must
be nearly quarter 
filled when the 
chemical potential
for the insertion 
of a hole is very
small by comparison
to the kinetic 
hopping scale 
$t$.

To conclude, in the final analysis,
we expect 
\begin{eqnarray}
\delta_{stripe} \approx 1/4,
\end{eqnarray}
where the approximation sign
signifies the relatively small
(compared to $t$) chemical 
potential shift due
to magnetic alleviation energies
and hole-hole interactions.

\section{Non Plaquette Pair States}

Here we show that unless the fundamental
pair building blocks are not chosen 
properly then the hole density may 
come out to be incorrect (when contrasted
with the very general single hole problem 
for consistency in the large t limit). 
The filling fraction will be larger than
a quarter found for the general
hole problem before assuming anything about
the possibility of pairing, phase separation
etc. amongst holes. We will now analyze diagonal pairs
which may be symmetrized and anti-symmetrized.
Apart from some trivial numerical
modifications, the considerations that 
we present here may be repeated
word for word for ``vertical'' pair states
extending along the rungs
or ``horizontal'' pairs extending
along the legs. There is a viable
link between the correct
symmetry of the pair state 
and its extension over different
non-overlapping plaquettes and
the anticipated hole density. In order
to exhibit this link and  
point to the viability 
of plaquette pair physics,
we will purposefully, focus on
a pair state ansatz 
that will give an 
incorrect filling
fraction in the $\mu/t =0$ limit. 
More complicated
and comprehensive treatments
may employ much of 
what is known about
the two leg ladders
(for instance, the
SO(8) symmetry present
in their low coupling 
limit \cite{balents}).

\subsection{Diagonal Pairs}

\begin{figure}
\includegraphics[width=8.2cm]{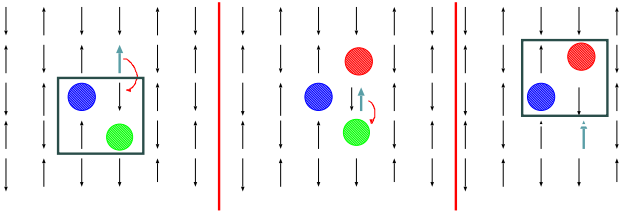}
\caption{A hole pair advancing along the 
stripe. If we want to account for
such motions (either as shown above,
or resulting from a higher order hopping 
process taking the pair in and
out of the antiferromagnet surrounding 
the stripe back to the stripe),
we must allow for overlapping plaquettes. As we will
see, including such possibilities
leads to higher filling 
fractions (i.e. $\delta \ge 1/3$
for a nearly vanishing relative 
chemical potential $(\mu/t)$).
}
\label{pairhopping}
\end{figure}

We will now write down an effective
one dimensional model for diagonal
pairs of two possible chiralities
(left/right tilting) as shown in
the right and left hand panels of
Fig.(\ref{pairhopping}). To give 
the reader a flavor of where
we are heading the general
argument may be summarized as 
follows- we will
effectively make an exact 
projection of the two 
rung ladder along its 
axis to produce 
an effective one dimensional
model.  Henceforth we will
denote the two natural diagonal pair 
chiralities (or polarizations) by a Greek index 
$\alpha = 1,2$ and further employ the 
shorthand

\begin{eqnarray}
\Delta_{i}^{(\alpha=1)}& = & c^{\dag }_{2, i, \sigma}c_{1,i+1,-\sigma }^{\dag }\nonumber \\
\Delta^{(\alpha =1 ) \dag }_{i} & = & c_{1,i+1,-\sigma}c_{2,i,\sigma}.
\label{pairdefn}
\end{eqnarray}
for the hole pair annihilation and creation operators
of the diagonal polarization of Fig.(\ref{pairhopping}) $(\alpha =1)$
in terms of the electronic operators.
In Eqn.(\ref{pairdefn}) the first subscript  
(which is always one or two) marks the location
of the hole on the ladder- on which of
the two rungs it is found (upper or lower) while
the second ($i$) denotes the location
of the hole along the stripe (ladder)
axis. The pair creation operator removes two electrons
and the pair annihilation operator creates two electrons.
Note that in these operators, there is no
spin index symmetrization/anti-symmetrization.
The spin polarizations are dictated
by the location along the ladder
(whether or not $i$ is even. 
We may conform 
to the convention that the spin subscript $\sigma = \uparrow$ if
$i$ is even and that $\sigma = \downarrow$ along
the odd rungs (i.e. we will choose
our origin $i=0$ in such a way
that the spin polarization
of both electrons on that 
rung is $\sigma =\uparrow$). 
The introduction of each 
diagonal pair states
removes two bad magnetic
bonds along the two
rungs that it occupies.

The number associated with 
these hole (``h'') pair operators is, as usual,
\begin{eqnarray}
n_{i,h}^{(\alpha)} =  
[\Delta^{(\alpha)}_{i}]^{\dag} \Delta^{(\alpha)}_{i}.
\end{eqnarray} 
We therefore can write the most general 
effective one dimensional Hamiltonian
\begin{eqnarray}
\label{H in terms of pair operators}
\tilde{H}_{eff}= - \sum _{ \langle ij \rangle} t_{\alpha \beta}^{ij}
(\Delta_{i}^{(\alpha) \dag
} \Delta_{j}^{(\beta)} + \Delta^{ (\beta) \dag 
}_{j}\Delta_{i}^{(\alpha)}) \nonumber
\\ + \sum_{\langle ij \rangle} V^{ij}_{\alpha \beta}
n_{i,h}^{(\alpha)}  n_{j,h }^{(\beta)}   + \tilde{J}^{\prime} \sum_{i, \alpha}
\Delta_{i}^{(\alpha) \dag }  \Delta_{i}^{(\alpha)} + ...
\label{general}
\end{eqnarray} 
where the ellipsis denote higher order terms in $\{\Delta_{i} \}$.
All potential terms appear here including 
ones with infinite long range interactions (e.g. Coulomb effects)
in which terms like $V_{ij}$ decay as slowly 
as desired as a function of the separation 
$|i-j|$. Note that as a single hole 
cannot be doubly occupied (or an
electron of fixed spin cannot be twice
removed), no two diagonal bonds
may share the same site; this
may formulated in terms 
of an effective infinite 
hard core repulsion: 
$V_{\alpha =1, \beta =2}^{i, i+1} = V_{2,1}^{i,i+1} = \infty$.

Employing the Matsubara-Matsuda transformation
once again we arrive at a flavored version of Eqn.(\ref{general})
where we dress the various spins by the polarization
(diagonal orientation of the pair that they represent).
Insofar as commutators 
are concerned, the two polarization
sectors are completely 
decoupled. Setting, for each $\alpha$, \( \Delta_{i}= S_{i}^{-} \)
and \( \Delta_{i}^{\dag }= S_{i}^{+} \) 
we see
that the pair creation- and annihilation-operators form an SU(2) spin
algebra. The effective Hamiltonian is a trivially
flavored version of Eqn.(\ref{sof2}):

\begin{eqnarray}
\! \! \! \! \! \! \! \! \! \! \! \! \tilde{H}_{eff}=
- \sum_{\langle ij \rangle} \tilde{J}^{\perp ij}_{\alpha \beta}
(S_{i}^{x (\alpha)} S_{j}^{x (\beta)}+S_{i}^{y (\alpha)}  S_{j}^{y
(\beta)}) \nonumber
\\ + \sum _{\langle ij \rangle} \tilde{J}^{z ij}_{\alpha \beta} 
S_{i}^{z (\alpha)} S_{j}^{z (\beta)} - \tilde{J}^{\prime} \sum_{i, \alpha}
(S_{i}^{z (\alpha)})^{2} \nonumber
\\ - \sum_{i, \alpha}  \tilde{h}
S_{i}^{z(\alpha)}  
+ ... 
\label{Heisenberg formula}
\end{eqnarray}
where the ellipsis denote higher order spin interactions.
As before, to quadratic order, comparison with Eqn.(\ref{H in terms of pair
operators}) shows 
that Eqn.(\ref{Heisenberg formula}) is nothing but an XXZ model 
with $\tilde{J}^{\perp i j}_{\alpha \beta} = 2 t_{\alpha \beta}^{ij}$,
$\tilde{J}^{z i j}_{\alpha \beta}= V^{ij}_{\alpha \beta}$ 
and $\tilde{h} = (2 V + J)$ (with $J$ the magnetic
alleviation energy),
if $V = V^{ij}_{\alpha \alpha}$ is constant 
for all nearest neighbor links $|i-j|=1$.
The only limitation of Eqs.(\ref{H in terms of pair
operators}, \ref{Heisenberg formula}) is that
they only allow for diagonal pairs as the 
fundamental building blocks (e.g. 
no single electron excitations
are accounted for).

\subsection{Complete Fock Space (Lorentz) Algebra}

With an eye to things to come
let reshuffle the two 
diagonal sectors $\alpha=1$
and $\alpha=2$ 
(corresponding to
the left and right most
real space 
configurations
of Fig.(\ref{pairhopping}).

By a simple linear transformation
\begin{eqnarray}
\vec{\tilde{S}} = \vec{S}^{(\alpha=1)} + \vec{S}^{(\alpha=2)}
\nonumber
\\ \vec{\tilde{A}} =  \vec{S}^{(\alpha=1)} - \vec{S}^{(\alpha=2)}
\label{Lorentz}
\end{eqnarray}
the $SU(2) \otimes SU(2)$ algebra of the two diagonal
sectors transforms into a Lorentz ($SO(4)$) 
algebra. Physically, 
\begin{eqnarray}
\tilde{S}^{\pm}_{i} = \Delta^{1}_{i} + \Delta^{2}_{i} \nonumber
\\ \tilde{A}_{\pm}^{i} = \Delta^{1}_{i}- \Delta^{2}_{i}
\label{genesis}
\end{eqnarray} 
are the creation operators 
for the symmetric and 
anti-symmetric rung pair states.
The symmetric combination, the angular 
momentum vector $\vec{S}$ (or $\vec{L}$), 
forms an $SO(3)$ subgroup: $[\tilde{S}^{i},\tilde{S}^{j}] = i \epsilon_{ijk} S^{k}$
with the indices $i,j$ and $k$ now denoting the standard spatial
components. 
The z-component
\begin{eqnarray}
\tilde{S}_{i}^{z} = 1  - \sum_{\alpha=1,2} n_{i,e}^{(\alpha)}
\label{S=1} 
\end{eqnarray}   
may assume the eigenvalues $\{1,0,-1\}$
corresponding to a $2 \times 2$ patch 
of the ladder being electronically full, 
having one diagonal hole pair, and two having diagonal hole pairs.
The symmetric combinations $\{\vec{\tilde{S}}_{i}\}$
satisfy $SO(3)$ algebra 
with spin one. The net number
of electrons is given by 
the total magnetization
of the symmetric $\langle \tilde{S}^{z}_{tot}\rangle$ 
component.

\subsection{Restricted Symmetric Hole Pair State Basis}

We will now make the physical assumption (fortified by our earlier 
analysis \cite{us}) 
that only the 
the low energy symmetric hole pair 
bonding states appear at low energies.
Defining, within the restricted symmetric hole pair state basis,   
\begin{eqnarray}
S^{-}_{i} = \frac{1}{\sqrt{2}} (\Delta_{i}^{1} + \Delta_{i}^{2})
\equiv \Delta_{i}, \nonumber
\\ S^{+}_{i} = \frac{1}{\sqrt{2}} (\Delta_{i}^{1 ~ \dagger} +
\Delta_{i}^{2 ~ \dagger}) \equiv \Delta_{i}^{\dagger},
\nonumber
\\ S_{i}^{z} = \frac{1}{2} [ 1- 2 n_{i; symm}],
\label{S_defn}
\end{eqnarray}
we see that $\vec{S}$ satisfies $SU(2)$ algebra with $S=1/2$
within the restricted symmetric bonding state space
$\prod_{i} (\Delta_{i}^{\dagger})^{r_{i}} | 0 \rangle$ with $r_{i} =0$
or 1, and where $|0 \rangle$ denotes the vacuum state in which
no holes are present. The bosonic hole pair states fill up the ladder 
as the electronic orbital states do the in 
atomic and molecular systems. We may regard this 
as an effective Hund's
rule. At low hole filling fractions, only the low energy 
bonding states will be occupied.
At high hole occupancies (in excess of
half a hole per unit site), anti-symmetric
anti-boding states will also appear and
the full $SO(4)$ algebra of
the full blown unrestricted Fock space 
will raise its head.

Substituting Eqn.(\ref{genesis},\ref{S=1},\ref{S_defn}) 
into Eqn.(\ref{general}) and omitting any terms
containing anti-bonding operators (which will
take us out of the restricted symmetric pair state
subspace) an arbitrarily high polynomial
in $\{\vec{S}_{i}\}$
will result identical
to Eqn.(\ref{sof2}).
The hard core repulsion term 
$V [n_{i}^{(\alpha=1)} n_{i+1}^{(\alpha=2)} + n_{i}^{(\alpha=2)}
n_{i+1}^{(\alpha=1)}]$ transforms into an analogous hard 
core term $V S_{i}^{z} S_{i+1}^{z}$ for the z-component
of the spins (the distance between 
up spins cannot be smaller than two).

Such a hard core term may be interpreted from 
first principles. A state containing
nearest neighbor symmetric bonding pairs
is not normalized to one. To have normalization
we must consider the correlated four hole
state $2^{-1/2} \sum_{\alpha=1,2} \Delta^{\alpha~ \dagger}_{i}
\Delta^{\alpha~ \dagger}_{i+1} | 0 \rangle$. The
reduced norm (probability) of the nearest neighbor
bonding pairs vis a vis other bonding 
pairs is similar to having a 
large potential barrier.

\subsection{One Third Doping And Beyond}

Sans the hard core constraint, the model
has its ground state at half filling
(i.e. the density of up spins 
is a half): Whenever an XXZ model has a nearest neighbor 
hard core constraint (whenever the distance between 
up spins cannot be smaller than two),
the density of up spins is equal 
to a third within the ground state
\cite{XXZ}. A careful counting of
the various diagonal pair states
reveals that the hole occupancy
within the stripe is expected
to be 1/3 in the large $t$ limit.
Repeating the same exercise 
for hole pairs along
rungs and legs we arrive
at similar large
hole densities.

\section{Single Electronic Description}
\label{single}

\subsection{Symmetric Rung States}

In an earlier work \cite{us}, we have numerically obtained
the single hole and hole pair energy spectrum on
a stripe. Let us first restrict attention to the
single electronic description. The states along the 
various rungs form resonant bonding states. 
We will now make the physical assumption (fortified by numerics) that only the 
the low energy symmetric (or anti-symmetric) hole rung 
bonding states appear at low energies. We will now consider
the symmetric (anti-symmetric) smearing of a hole
along the two legs of the rung.
Defining, within the restricted symmetric hole rung state basis,   
the two operators
\begin{eqnarray}
c_{i;\pm} \equiv \frac{1}{\sqrt{2}} [c^{(\alpha =1)}_{i} \pm c^{(\alpha=2)}_{i}],
\end{eqnarray}
with $c$ the electronic annihilation
operator, $\alpha = 1,2$ now a leg index, 
and $i$ a rung label, it is readily verified
that these operators satisfy disjoint 
canonical Fermi anti commutation relations:
$\{c_{i; \pm}, c_{j,\pm}^{+}\} = \delta_{ij}$.

We will assume that the 
even parity symmetric bonding 
hole states are lower in 
energy and restrict ourselves to 
that basis. In the problem
of physical relevance,
$|t_{i=j}^{\alpha=1,\beta=2}| \gg $ all other hopping
amplitudes and the hole will quickly resonate between the two rungs
of the ladder before inching its way along the 
ladder axis.


Once again,
the hole states fill up the ladder 
as the electronic orbital states do the in 
atomic and molecular systems. We may regard this 
as an effective Hund
rule. At low hole filling fractions, only the low energy 
bonding states will be occupied.
At high hole occupancies (in excess of
half a hole per rung of a bond 
centered stripe or a hole
per plaquette in the cuprate
plane), anti-symmetric
anti-boding states will also appear and
the algebra of
the full blown unrestricted Fock space 
will raise its head.

In what will follow shortly,
we will treat the symmetric
rung holes as spinless
fermions. An average 
density of half a spinless fermion
(the particle-hole symmetric
point) per rung corresponds to
a quarter hole per site.

We will consider the
extended Hubbard model in the 
restricted basis of these 
low lying symmetric
rung states.

\subsection{The Ideal Spinless Fermi Gas}
\label{ideal}

As we review in Appendix(\ref{any}), within the general Hamiltonian
Eqn.(\ref{Hubbard model+}) in the limit of infinite on-site
repulsion $U$, the charge degrees
of freedom transform into those of a spinless
Fermi system. In the bare Hubbard model, 
at $U= \infty$ the system reduces exactly 
an ideal (non-interacting) Fermi gas
with a dispersion 
\begin{eqnarray}
\epsilon_{k} = -2t \cos k.
\end{eqnarray}

For a non-interacting Fermi gas, the zero temperature
occupancy
\begin{eqnarray}
\langle n_{k} \rangle = \theta(\mu -\epsilon_{k}).
\end{eqnarray}

If inserting a hole leads to no change
in the energy balance then the chemical
potential, by its very definition, vanishes.
The energetics of the chemical potential
acts as a Lagrange multiplier
enforcing a certain average occupancy.
The energy of adding or removing an electron
is the same (zero) at a chemical potential 
$\mu =0$.  By particle-hole
symmetry, at $\mu =0$ the chemical potential lies 
in the middle of the band and the 
occupancy of the rung symmetrized spinless fermi
particles is a half- and that
of the holes is 1/4. The particle-hole
symmetry of the kinetic
energy band is dictated by 
the hermiticity of the Hamiltonian.
In fact, even if the Hamiltonian included
arbitrary long range hoppings $\{t_{r}\}$
and reduced to the general form 
\begin{eqnarray}
H = - \sum_{ij} t_{|j-i|} (c_{i}^{\dagger} c_{j} + h.c.),
\end{eqnarray}
in the infinite $U$ limit, then by hermiticity
whenever the chemical potential is zero the Hamiltonian
will be particle hole symmetric. In that case, the 
density of rung symmetrized spinless fermions 
will be a half, and the
density of holes along the bond centered stripe 
will be 1/4.

If the addition of holes (or removal of 
electrons) leads to 
a reduction in the magnetic strain energy
along the stripe then the chemical
potential for the electrons is reduced
(or, equivalently, that for holes,
is increased).
A removal of a bad bond along the
rung leads to a lowering of the 
magnetic energy by $J$. Thus 
the hole chemical potential
$\mu = J$ and the net, 
rung symmetrized, charge 
occupancy is given by an integral of 
$\langle n_{k} \rangle$. The hole
density is $\langle (n_{k}/2) \rangle$.
This expectation value (hole occupancy) 
changes slightly with temperature
according to the evolution of the 
Fermi function.

It is worthwhile viewing this, one last time,
in terms of unbroken spin rotational 
symmetry in an effective XX(Z) model.

We now apply the Jordan-Wigner transformation
in order to arrive at XX(Z) model. As well known, 
the infinite $U$ Hubbard Hamiltonian trivially transforms
into  
\begin{eqnarray}
H = - 2t \sum_{i} (S_{i,x} S_{i+1,x} + 
S_{i,y} S_{i+1,y}).
\label{XY}
\end{eqnarray}
An infinitesimal chemical potential leads to
an additional small magnetic field coupling
to $\sigma_{z}$. The absence of 
spin symmetry breaking in this trivial
example (no $\sigma_{z}$ interaction
appears in $H$) leads to the conclusion
that the band is half filled (or that the 
hole density is 1/4).

This is immediately seen taking
note of  
\begin{eqnarray}
S_{i}^{z} = 1/2 - c_{i}^{\dagger} c_{i}.
\end{eqnarray}

A vanishing $\langle \sigma_{z} \rangle$ for
the XY Hamiltonian of Eqn.(\ref{XY}) leads to
$\langle n_{j} \rangle = 1/2$. Related back
to our original problem, this implies
a quarter empty electronic bond centered 
stripe. This value is seen
here to be dictated by 
an unbroken particle-hole
symmetry.

If additional interactions next nearest neighbor
pieces of the Coulomb and other interactions ($\sum_{\langle ij \rangle}  
V_{ij} n_{i} n_{j}$) and the magnetic energy
alleviation energy $J$ are added to the Hamiltonian
then the Hamiltonian for the spinless symmetrized 
holes readily transforms to a full blown XXZ model with

\begin{eqnarray}
H = - 2t \sum_{i} (S_{ix} S_{i+1,x} + 
S_{iy} S_{i+1,y}) \nonumber
\\ + \sum_{\langle i j \rangle} 
\frac{V_{ij}}{4}  (1- S_{i}^{z}/2) (1- S_{j}^{z}/2) 
- \frac{J}{2} \sum_{i} (1-S_{i}^{z}/2)
\nonumber
\\ \equiv \sum_{\langle i j \rangle} [-J_{\perp} (S^{+} S^{-}
+ h.c.) + J_{z} S_{i}^{z} S_{j}^{z}] \nonumber
\\ + \sum_{i} h S_{i}^{z}.
\end{eqnarray}

We first note that if the electrostatic
energy $V_{i, i+1}$ is such that it equals
the magnetic alleviation energy $J$ 
then by the absence of spontaneous symmetry
breaking in one dimension, $\langle S^{z} \rangle =0$
and the bond centered stripe is quarter
filled. If, hypothetically, $S^{z}$ would spontaneously
develop a nonzero magnetization then this
would imply that even if in an ensemble of stripes,
the overall density of holes would be a quarter 
on average, different stripes in the ensemble 
would exhibit two different densities about 
that mean. 

An imbalance between $V$ and $J$ leads to nonzero 
$h$ and to a finite 
value of $\langle S^{z} \rangle \neq 0$-
a deviation from quarter filling on every stripe. 
In the above we took into account both the Coulomb and
magnetic alleviation effects in one 
go. 

In a very nice paper by Nayak and Wilczek \cite{nayak}
it was observed that 1/4 filling of a doped chain 
is indeed predicted 
from the Bethe ansatz solution of Lieb and Wu \cite{lieb}.
Some of the symmetries of the 1/4 filled
point were noted. Here we emphasize that at low chemical
potential, quarter filling
not only coincides with symmetries, 
but is, in fact, a rigorous outcome of
symmetry considerations. Here we also
note that a finite
magnetic alleviation energy
of the holes amounts to a 
shift in the chemical potential. 
The approximate calculations employed
earlier are not mandatory
for the determination of the 
hole occupancy within the 
ground state; in order
to account for magnetic
effects, we may simply set the hole chemical
potential $\mu = J$, and subsequently 
integrate the known ideal spinless 
fermi number density $\langle n_{\epsilon} \rangle$
up to that energy to obtain an exact result.

\appendix
\section{A Review of Trivial Spin Charge Separation 
in Any Dimension in the
Limit of Large On site Repulsion}
\label{any}

Here we review trivial spin-charge separation
within any Doped Mott insulator (of arbitrary
dimensions) in the limit of large on site 
repulsion energy $U \to \infty$. Let
$\psi_{\sigma_{1},...\sigma_{N}}(x_{1},...,x_{N})$ be an electronic 
or ``hole'' eigenstate of the general Hubbard Hamiltonian 
augmented by all possible 
higher order interactions (these may be sparked by electron phonon
terms, Coulomb repulsions etc.) and all possible range hopping amplitudes
$t,t^{\prime}, t^{\prime \prime}, ...$ etc.

Recall that, trivially, $\psi_{\sigma_{1},...\sigma_{N}}(x_{1},...,x_{N})$,

$\bullet$ is anti-symmetric by virtue of its electronic 
constituents: $\psi_{\sigma_{i} \sigma_{j}}(x_{i},x_{j}) 
= - \psi_{\sigma_{j} \sigma_{i}} (x_{j},x_{i})$.

Note furthermore that exactly at the $U \to \infty$ limit,

$\bullet$ $\psi(x_{i}=x_{j}) = 0$ irrespective of the spin indices
$\sigma_{1}, ..., \sigma_{N}$. The reason for this nodal behavior
is trivial- opposite spin occupation is forbidden by a divergent
on site penalty $U$, and parallel spin occupation is strictly 
forbidden by the Pauli principle. Also, at finite temperature
the probability for having any two 
particles occupy the same state
strictly vanishes.

If we invoke this nodal, hard core, condition, then we may now remove 
the on site Hubbard repulsion from the original Hubbard
Hamiltonian to obtain a general translationally invariant
spin independent Hamiltonian H. Herein lies the crux of
the trivial spin charge separation at infinite $U$. 

The most general solution to the Schrodinger equation 
$H|\psi \rangle = E |\psi \rangle$  satisfying 
the last nodal condition with a
spin independent Hamiltonian $H$ 
(all terms apart from the on site interaction
U are spin independent) is 
\begin{eqnarray}
\psi_{\sigma_{1},...,\sigma_{N}}(x_{1},...,x_{N}) =
\chi(\sigma_{1},...,\sigma_{N})  W(x_{1},...,x_{N}),
\end{eqnarray}
and linear superpositions of such degenerate solutions,  
where $W$ is a solution to the Schrodinger equation.
Here the trivial spin-charge separation is manifest.

Let us now impose the Fermionic statistics condition
\begin{eqnarray}
P^{ij}_{spin}P^{ij}_{charge} = -1
\end{eqnarray}
for all $i \neq j$. Note that $[P^{ij}_{spin},H]= 0$ as the
Hamiltonian
is spin independent. As
$P^{ij}_{charge} = -[P^{ij}_{spin}]^{-1} = - P^{ij}_{spin}$, the charge
permutation operator also commutes with $H$. This implies
the function $W$ 
must satisfy 
\begin{eqnarray}
P^{ij}_{charge} W(x_{1},... ,x_{i}, ..., x_{j}, ..., x_{N}) = \nonumber
\\ 
\alpha_{charge}^{ij} W( x_{1},... ,x_{j}, ..., x_{i}, ..., x_{N}).
\end{eqnarray}

The identity $[P^{ij}_{charge}]^{2} =1$ 
implies that $\alpha^{ij}_{charge} =\pm 1$.

In general, the energy of a nodeless function
$W$ is expected to be lower. However, the charge degrees 
of freedom cannot be completely symmetric: if the spinor 
$\chi(\sigma_{1},...,\sigma_{N})$
is fermionic (completely anti-symmetric) then it must vanish identically
for $N\ge 3$ particles for the spin $S=1/2$ electrons. 
In any dimension, the effective Hamiltonian
can now be written in terms of the 
charge only spinless degrees of freedom.
If the Hamiltonian is particle-hole
symmetric (as expected by hermiticity of
the most general arbitrary range kinetic 
terms) then whenever the chemical 
potential vanishes (whenever the insertion 
of a hole and a particle 
both cost zero 
energy), the average ensemble density 
of the correct spinless charge degrees of 
freedom must be a half. 
If, for instance, {\em in any dimension}, all low energy physics
could be captured in terms of 
spinless charge pair degrees
of freedom $W = F(\{\phi(x_{i},x_{j})\})$
then in terms of these pairs, the 
ground state of the system would
be half occupied. 

The charge degrees of freedom encapsulated 
in $W$ can be symmetric with respect to, at most,
single pairs (and anti-symmetric
within these pairs); a higher symmetry 
is ruled out by the impossibility of a spinor
anti-symmetric in three and more spin 1/2
indices (spinons cannot be fermionic). 

Let us now regress to one dimension.
Here, hard core bosons cannot be
distinguished from spinless fermions and different symmetry
states of $W$ are in fact degenerate at the 
$U = \infty$ point. This degeneracy is lifted
as $t/U$ becomes arbitrarily small but finite. As $U$ 
is extremely large but finite $0 = |\psi_{\sigma_{i} = \sigma_{j}}
(x_{i}=x_{j})|< |\psi_{\sigma_{i} =
-\sigma_{j}}(x_{i}=x_{j})| \ll 1$ and local 
singlet correlations are generated.

It has indeed been established by one dimensional 
Bethe ansatz \cite{Ogata} that at infinite $U$, the 
function $W$ is a Slater determinant 
of single particle momentum 
eigenstates.

\begin{acknowledgments}

The authors would like to thank Wim van Saarloos and Jan Zaanen for helpful remarks and encouragement.
\end{acknowledgments}

\end{document}